\documentclass[epj]{svjour}
\usepackage{graphics}
\usepackage{amsmath}
\DeclareMathOperator\arctanh{arctanh}

\newcommand{\be}{\begin{equation}}
\newcommand{\ee}{\end{equation}}
\newcommand{\bea}{\begin{eqnarray}}
\newcommand{\eea}{\end{eqnarray}}

\begin{document}
\title{Doubleverse entanglement in third quantized non-minimally coupled varying constants cosmologies}

\titlerunning{Doubleverse entanglement in third quantized varying constants cosmologies}


\author{Adam Balcerzak\inst{1,2} \and Konrad Marosek\inst{3}
}

\authorrunning{A. Balcerzak, K. Marosek}
%
%
\institute{Institute of Physics, University of Szczecin,
Wielkopolska 15, 70-451 Szczecin,  Poland \and Copernicus Center for Interdisciplinary Studies, Szczepa\'nska 1/5, 31-011 Krak\'ow, Poland \and  Chair of Physics, Maritime University of Szczecin, Wa{\l }y Chrobrego 1-2, 70-500 Szczecin, Poland}
\date{Received: date / Revised version: date}
%
\abstract{In this paper we consider a third quantized cosmological model with varying speed of light $c$ and varying gravitational constant $G$ both represented by non-minimally coupled scalar fields. The third quantization of such a model leads to a scenario of the doubleverse with the two components being quantum mechanically entangled. We calculate the two parameters describing the entanglement, namely: the energy and the entropy of entanglement where the latter appears to be a proper measure of the entanglement. We consider a possibility that the entanglement can manifests itself as an effective perfect fluid characterized by the time dependent barotropic index $w_{eff}$, which for some specific case corresponds to the fluid of cosmic strings. It seems that such an entanglement induced effective perfect fluid may generate significant backreaction effect at early times.
\PACS{
      {04.50.Kd}{Modified theories of gravity}   \and
      {04.60.−m}{Quantum gravity}
     } 
} 
\maketitle

\section{Introduction}
\label{intro}
The idea of multiverse assumes that our universe is a part of a larger whole - a multiverse being a collection of many universes. The four different types of the relation between our universe and the rest of the multiverse were defined \cite{tegmark}. The most obvious type of the relation assumes that the rest of multiverse is the space outside the observationally accessible region (level I multiverse). The one more elaborated defines our universe as one of the causally disconnected post-inflationary bubbles with possibly different values of the physical constants (level II multiverse). The other two types involve the idea of Everett's many-worlds interpretation of quantum mechanics (level III multiverse) or treating large well defined purely mathematical structures as the existing elements of the multiverse (level IV multiverse). An interesting case (level II and III) defining the paradigm of interacting universes describes the interaction between the universes as occurring in the minisuperspace via quadratic terms \cite{Serrano,Robles1,Bertolami}. The causal disconnection present in level II multiverse in such models can be maintained. Another approach realising the level I multiverse investigates the effects of the entanglement between different possibly causally disconnected patches of the universe \cite{Holman1,Holman2}. An extraordinary approach to the concept of multiverse defined in \cite{Robles_ent1,Robles_ent2} is based on the so-called third quantization procedure which exploits the formal analogy between the Wheeler-DeWitt and the Klein-Gordon equations. In this approach the Klein-Gordon field is substituted by the wave function which is promoted in the course of the third quantization to be an operator acting on the Hilbert space spanned by the orthonormal set of vectors representing occupation with universes characterized by appropriate quantum numbers. A great advantage of this approach is that it naturally introduces quantum entanglement between universes and
provides tools to describe an interuniversal entanglement in terms of the thermodynamical quantities \cite{Robles_ent1,Robles_ent2,Robles_Bal,Alicki}. However, the connection between the ordinary thermodynamics and the thermodynamics of quantum entanglement is still not well understood.\\
\indent Many different cosmological scenarios have been considered so far in the context of the third quantization. We mention here an embedding of Brans-Dicke gravity in the third quantization scheme which interestingly leads to scenarios in which whole multiverse is created out of vacuum \cite{Pimentel}, an application of third quantization procedure to the varying constants model \cite{Balcerzak2} with non-minimally coupled dynamical scalar fields representing the speed of light and the gravitational constant \cite{Balcerzak1} which results in similar scenario of the multiverse creation or eventually the third quantization of the varying gravitational constant cyclic scenarios \cite{Marosek} in which the naturally arisen interuniversal entanglement leads to interesting behavior of the thermodynamical quantities \cite {Robles_Bal}. The third quantization procedure was also used to discuss
the transition from expanding to contracting cosmological phase (and vice-versa) in \cite{Buonanno,Gasperini}.\\
\indent
Our paper is organized as follows. In Sec. \ref{sec:1} we introduce the non-minimally coupled varying speed of light $c$ and varying gravitational constant $G$ theory defined in \cite{Balcerzak1} and describe the procedure of the third quantization of such a theory. In Sec. \ref{sec:2} we discuss based on the theory described in Sec. \ref{sec:1} a scenario which results in the emergence of the entanglement in a pair of universes which initially were disentangled. 
We also calculate the reduced density matrix of a single universe as well as its eigenvalues. In Sec. \ref{sec:3} we calculate the energy and the entropy of entanglement where the latter appears to be a proper measure of entanglement. In Sec. \ref{sec:4} we relate the previously calculated energy of entanglement with the classical energy-momentum content of the universe and argue that the entanglement can effectively simulate a perfect fluid with time dependent barotropic index. In Sec. \ref{sec:conc} we give our conclusions.

\section{Third quantized non-minimally coupled varying constants cosmological model}
\label{sec:1}
Our considerations are based on the model defined in \cite{Balcerzak2,Balcerzak1} which describes the variation of the speed of light and the variation of the gravitational constant with both quantities  represented by the two non-minimally coupled scalar fields. Such a model was originally inspired by the covariant and locally Lorentz-invariant varying speed of light theories \cite{Magueijo} and is given by the following action:
\begin{equation}
\label{action}
S=\int \sqrt{-g}  \left(\frac{e^{\phi}}{e^{\psi}}\right) \left[R+\Lambda + \omega (\partial_\mu \phi \partial^\mu \phi + \partial_\mu \psi \partial^\mu \psi)\right] d^4x,
\end{equation}
where $\phi$ and $\psi$ are some non-minimally coupled scalar fields, $R$ is the Ricci scalar, $\Lambda$ plays the role of the cosmological constant and $\omega$ is some parameter of the model. The action (\ref{action}) was obtained by replacing the speed of light $c$ and the gravitational constant $G$ in the original Einstein-Hilbert action with certain functions of the scalar degrees of freedom $\phi$ and $\psi$. The specific form of the relationship between the scalar fields $\phi$ and $\psi$ and the fundamental constants $c$ and $G$ is given by the following formulas:
\bea
c^3&=&e^{\phi}, \\
G&=&e^\psi.
\eea
This way the evolution of $\phi$ and $\psi$ by definition determine the variability of $c$ and $G$. By application of the field redefinition given by
\begin{eqnarray}
\label{fred}
\phi &=& \frac{\beta}{\sqrt{2\omega}}+\frac{1}{2} \ln \delta, \\
\psi &=& \frac{\beta}{\sqrt{2\omega}}-\frac{1}{2} \ln \delta,
\end{eqnarray}
the action (\ref{action}) can be rewritten in the form of the Brans-Dicke action which reads:
\bea
\label{actionBD}
S=\int \sqrt{-g}\left[ \delta (R+\Lambda) +\frac{\omega}{2}\frac{\partial_\mu \delta \partial^\mu \delta}{\delta} + \delta \partial_\mu \beta \partial^\mu \beta\right]d^4x.
\eea
The dependence of $c$ on space-time coordinates breaks the general covariance of the theory and entails specification of a coordinate system in which the theory of varying $c$ and $G$ is described by the action given by (\ref{action}) or (\ref{actionBD}). In other words our model needs to be formulated in a preferred reference frame. In fact this is a generic feature of the large class of the theories which deal with the problem of varying speed of light \cite{Magueijo}. Following the suggestions given in \cite{Magueijo} we will associate the preffered frame to formulate our model with the cosmological frame defined by flat FLRW metric given by:
\be
\label{FLRW}
ds^2=-N^2(dx^0)^2+ a^2(dr^2 + r^2 d\Omega^2),
\ee
where $N$ is the lapse function while $a$ is the scale factor both depending on coordinate $x^0$. The action (\ref{actionBD}) in the cosmological frame defined by the metric (\ref{FLRW}) takes the following form:
\begin{eqnarray}
\label{action_sym} \nonumber
S &=& \frac{3 V_0}{8 \pi} \int dx^0 \left(-\frac{a^2}{N} a' \delta' - \frac{\delta}{N} a a'^2   + \Lambda \delta a^3 N  \right. \\
&-& \left.\frac{\omega}{2} \frac{a^3}{N} \frac{\delta'^2}{\delta}-\frac{a^3}{N}\delta \beta'^2 \right),
\end{eqnarray}
where $()'\equiv \frac{\partial}{\partial x^0}$. In the gauge given by
\be
\label{gauge}
N=a^3\delta,
\ee
the solution of the model defined by action (\ref{actionBD}) is \cite{Balcerzak1}:
\bea
\label{rozwio1}
a&=& \frac{1}{D^2 {(e^{ F x^0})}^2 \sinh ^ M |\sqrt{(A^2-9)\Lambda }x^0| },\\
\label{rozwio2}
\delta &=& \frac{D^6 {(e^{ F x^0})}^6}{\sinh ^ W |\sqrt{(A^2-9)\Lambda }x^0|},
\eea
where   $A=\frac{1}{\sqrt{1-2\omega}}$, $M=\frac{3-A^2}{9-A^2}$, $W=\frac{2A^2}{9-A^2}$ and $D$ and $F$ are some integration constants. Due to the particular choice of the gauge (\ref{gauge}) the variable $x^0$ cannot be interpreted as the cosmic time and the relationship between the two variables can be retrieved be first finding the relation between $x^0$ and the rescaled cosmic time $\bar{x}^0$ defined by
\be
\label{res}
d\bar{x}^0\equiv N dx^0 = a^3\delta dx^0,
\ee
and then by solving for the proper time encountered by the comoving observer $\tau$ (the usual cosmic time) from the following formula:
\be
\label{prop}
d\tau \equiv \frac{|ds_{com}|}{c(\bar{x}^0)}= \frac{d\bar{x}^0}{c(\bar{x}^0)},
\ee
where $ds_{com}$ is the line element (\ref{FLRW}) evaluated on the world line of the comoving observer.
The formula (\ref{prop}) encodes the typical impact of varying speed of light on classical trajectories due to explicit dependence of the metric on the speed of light $c$ (see \cite{Magueijo}). Inserting (\ref{rozwio1}) and (\ref{rozwio2}) into (\ref{res}) and then integrating (\ref{res}) leads to the following relation between $x^0$ and the rescaled cosmic time $\bar{x}^0$ \cite{Balcerzak1}:
\begin{equation}
\label{conect}
\begin{split}
x^0 &= \frac{2}{\sqrt{(A^2-9)\Lambda}}  \arctanh \left(e^{\sqrt{(A^2-9)\Lambda}\bar{x}^0}\right)\,,
\hspace{0.2cm}
\text{for $\bar{x}^0<0$}\,,
\\
x^0 &= \frac{2}{\sqrt{(A^2-9)\Lambda}}  \arctanh \left(e^{- \sqrt{(A^2-9)\Lambda}\bar{x}^0}\right)\,,
\hspace{0.2cm}
\text{for $\bar{x}^0>0$}\,,
\end{split}
\end{equation}
where as in \cite{Balcerzak1} we will limit our considerations to the cases with $A^2>9$. The solution given by (\ref{rozwio1}) and (\ref{rozwio2}) together with (\ref{conect}) describes the pre-big-bang contraction that takes place for $\bar{x}^0<0$ followed by the post-big-bang expansion which occurs for $\bar{x}^0>0$. Both phases are separated by the curvature singularity which occurs for $\bar{x}^0=0$.
\begin{figure}
\begin{center}
\resizebox{0.4\textwidth}{!}{\includegraphics{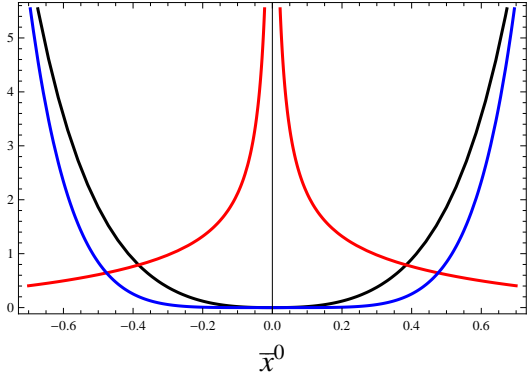}}
\caption{\label{acg} Qualitative behavior of the scale factor $a$ (black), the speed of light $c$ (red) and the gravitational constant $G$ (blue) before ($\bar{x}^0<0$) and after ($\bar{x}^0>0$) the curvature singularity plotted with the help of solution given by (\ref{rozwio1}), (\ref{rozwio2}) and (\ref{conect}).}
\end{center}
\end{figure}
Formulas (\ref{rozwio1}) and (\ref{rozwio2}) also include the information on the evolution of the fundamental constants $c$ and $G$. It turns out that the gravitational constant $G$ vanishes while the speed of light $c$ diverges as the universe approaches the curvature singularity at $\bar{x}^0=0$ (see Fig. (\ref{acg})).

We are also interested in the hamiltonian picture of the presented model. In order to find the corresponding hamiltonian we first observe that the action (\ref{action_sym}) in the new variables $\eta$, $x_1$ and $x_2$ defined by the following field transformations:
\bea
X = \ln(a \sqrt{\delta}),&& \hspace {0.3cm} Y = \frac{1}{2A} \ln \delta,\\
\eta=r(AY-3X), \hspace {0.1cm} x_1&=&r(3Y-AX), \hspace {0.1cm} x_2=2\sqrt{\tilde{V}_0}\beta,
\eea
where $\tilde{V}_0 = \frac{3 V_0}{8\pi}$ and $r=2\sqrt{\frac{\tilde{V}_0}{A^2-9}}$ simplifies to the following form:
\be
\label{action_simple}
S= \int dx^0 \left[\frac{1}{4}(\eta'^2-x_1'^2-x_2'^2)+\bar{\Lambda}e^{-2\frac{\eta}{r}}\right],
\ee
where $\bar{\Lambda} = \tilde{V}_0\Lambda$. The corresponding hamiltonian reads:
\be
\label{ham}
H=\pi_\eta^2 -\pi_{x_1}^2- \pi_{x_2}^2-\bar{\Lambda}e^{-2\frac{\eta}{r}},
\ee
where $\pi_\eta=\frac{\eta'}{2}$, $\pi_{x_1}=-\frac{x_1'}{2}$ and $\pi_{x_2}=-\frac{x_2'}{2}$ are the conjugated momenta. The form of the hamiltonian (\ref{ham}) suggests that both $\pi_{x_1}$ and $\pi_{x_2}$ are conserved during the evolution. This means the classical evolution is formally equivalent to the scattering of a particle on the exponential potential barrier. The solutions of the set of Hamilton equations corresponding to the hamiltonian (\ref{ham}) are:
\bea
\label{ham_sol1}
\eta&=&\ln \sinh|\sqrt{(A^2-9)\Lambda }x^0|, \\
\label{ham_sol2}
x_1&=& -2 \pi_{x_1} x^0 + E, \\
\label{ham_sol3}
x_2&=&-2 \pi_{x_2} x^0 + P,
\eea
where $E$ and $P$ are some integration constants. By examining the solution (\ref{ham_sol1}) we see that $\eta$ can define two regimes - the high-curvature regime (defined by the vanishing scale factor $a\rightarrow 0$) which corresponds to $\eta\rightarrow \infty$ and the low-curvature regime (defined by higher values of the scale factor $a$) which occurs for $\eta\rightarrow -\infty$. On the other hand it can be checked that the high-curvature regime (for $\eta\rightarrow \infty$) is characterized by the following asymptotic values of the momentum $\pi_\eta$:
\[ \pi_\eta = \left\{
  \begin{array}{l l}
    \sqrt{\bar{\Lambda}} & \quad \text{collapsing pre-big-bang solution}\\
    -\sqrt{\bar{\Lambda}} & \quad \text{expanding post-big-bang solution,}
  \end{array} \right.\]
while in the low-curvature regime (for $\eta\rightarrow -\infty$) we have that:
   \[ \pi_\eta = \left\{
  \begin{array}{l l}
    \sqrt{\bar{\Lambda}} e^{-\frac{\eta}{r}} & \quad \text{collapsing pre-big-bang solution}\\
    -\sqrt{\bar{\Lambda}} e^{-\frac{\eta}{r}} & \quad \text{expanding post-big-bang solution.}
  \end{array} \right.\]

In order to obtain the Wheeler-DeWitt equation which describes the quantum mechanical regime corresponding to the considered model we apply the Jordan quantization rules and replace the canonical momenta with the operators: $\pi_\eta\rightarrow \hat{\pi}_\eta=-i \frac{\partial}{\partial \eta}$, $\pi_{x_1}\rightarrow \hat{\pi}_{x_1}=-i \frac{\partial}{\partial x_1}$ and $\pi_{x_2}\rightarrow \hat{\pi}_{x_2}= -i \frac{\partial}{\partial x_2}$. The resulting Wheeler-DeWitt equation reads:
\begin{equation}
\label{KG}
\ddot{\Phi} - \Delta \Phi + m_{eff}^2(\eta) \Phi=0,
\end{equation}
where $\dot{( )}\equiv\frac{\partial}{\partial \eta}$, $\Delta = \frac{\partial^2}{\partial x_1^2}+\frac{\partial^2}{\partial x_2^2}$ and $m_{eff}^2(\eta)= \bar{\Lambda} e^{-\frac{2}{r}\eta}$.

The formal analogy between (\ref{KG}) and the Klein-Gordon equation allows us to perform the so-called third quantization procedure by formally applying the Klein-Gordon field quantization rules. It is assumed that the resulting theory involves the Fock space associated with the considered model of the multiverse.

The third quantized action that leads to the Wheeler-DeWitt equation given by (\ref{KG}) is:
\begin{eqnarray}
\label{3action}
S_{3Q}=\frac{1}{2}\int\left[ \dot{\Phi}^2 - (\nabla \Phi)^2 -m_{eff}^2(\eta) \Phi^2 \right]d^2x d\eta,
\end{eqnarray}
where $\nabla$ is a two-dimensional gradient operator associated with the the free degrees of freedom $x_1$ and $x_2$. The third quantized hamiltonian corresponding to the action (\ref{3action}) is:
\begin{eqnarray}
\label{3ham}
H_{3Q}(\eta)=\frac{1}{2}\int\left[ \pi^2 + (\nabla \Phi)^2 + m_{eff}^2(\eta) \Phi^2 \right]d^2x,
\end{eqnarray}
where the conjugated momentum $\pi=\dot{\Phi}$. A crucial step in the third quantization procedure involves choosing the vacuum. Generally the vacuum and the series of excited states of the wave function $\Phi$ associated with the choosen vacuum is determined by a set of particular mode functions $v_k(\eta)$ that are included in the usual expansion formula of the field operator $\hat{\Phi}$ given by \cite{Mukhanov}:
\begin{equation}
\label{exp1}
\hat{\Phi}(\vec{x},\eta)=\frac{1}{\sqrt{2}}\int \frac{d^2k}{2\pi}[e^{i\vec{k}\cdot\vec{x}}v_k^*(\eta)\hat{a}^-_{\vec{k}}+e^{-i\vec{k}\cdot\vec{x}}v_k(\eta)\hat{a}^+_{\vec{k}}],
\end{equation}
where $\vec{k}\equiv(k_1,k_2)$, $d^2k\equiv dk_1 dk_2$ and $|\vec{k}|\equiv k \equiv \sqrt{k_1^2+k_2^2}$. The mode functions $v_k(\eta)$ fulfill the following mode equation (a condition imposed by (\ref{KG})):
\be \label{modeeq}
v_k(\eta)''+\omega_k(\eta)^2 v_k(\eta)=0,
\ee
where $\omega_k(\eta)=\sqrt{k^2+m_{eff}^2(\eta)}$ and the normalization condition:
\begin{equation}
\label{normcon}
W(v_k(\eta),v^*_k(\eta))=2i,
\end{equation}
where $W(\cdot,\cdot)$ denotes wronskian. The creation and annihilation operators $\hat{a}^+_{\vec{k}}$ and $\hat{a}^-_{\vec{k}}$ that defines the ladder of the excited states of the field operator  $\hat{\Phi}$ fulfill the standard commutation relations:
\begin{eqnarray}
\label{commut}
[\hat{a}^-_{\vec{k}},\hat{a}^+_{\vec{k}'}]&=&\delta(\vec{k}-\vec{k}'), \\
\lbrack \hat{a}^-_{\vec{k}},\hat{a}^-_{\vec{k}'}\rbrack &=& 0,\\
\lbrack \hat{a}^+_{\vec{k}},\hat{a}^+_{\vec{k}'}\rbrack &=& 0.
\end{eqnarray}
By definition a vacuum state $|0\rangle$ is given by the usual condition:
\be
\label{vacst}
\hat{a}^-_{\vec{k}} |0\rangle=0
\ee
for all $\vec{k}$.
Naturally, the vacuum state $|0\rangle$ is not unique since it relies on the solution of the mode equation (\ref{modeeq}), which also cannot be uniquely specified.

\section{Emergence of entanglement in pairs of the universes}
\label{sec:2}

In this section we will show that the third quantized varying constants theories described in the previous section naturally involve scenarios in which the entanglement develops in previously disentangled pair of universes. 
A specific scenario can be implemented by appropriate selection of boundary conditions determining the initial state of the field operator $\hat{\Phi}$. It also requires selecting the vacuum. Since our model naturally defines the two asymptotic regions in the minisuperspace - the high-curvature one defined by vanishing of the scale factor $a$ which occurs for $\eta\rightarrow \infty$ (point $\bar{x}^0=0$ in Fig. (\ref{acg})) and the low-curvature one defined by higher values of the scale factor $a$ which appears for  $\eta\rightarrow -\infty$ - the selected vacua  will be associated with these two asymptotic regions. Specifically, we will define the high-curvature vacuum $|_{(in)}0\rangle$ (in-vacuum) as determined by the solutions of the mode equation (\ref{modeeq}) for $\eta\rightarrow\infty$ region of the minisuperspace and the low-curvature vacuum $|_{(out)}0\rangle$ (out-vacuum) as determined by the solutions of (\ref{modeeq})  for  $\eta\rightarrow-\infty$ region of the minisuperspace. We will be also assuming that the vacuum at any moment of the background evolution given by particular value of $\eta$ is controlled by  the instantaneous value of the background curvature.  In other words, the vacuum evolves along with the curvature and changes from the high-curvature in-vacuum $|_{(in)}0\rangle$ into the low-curvature out-vacuum $|_{(out)}0\rangle$ as the system moves between the two previously defined asymptotic regions of the minisuperspace (the high- and the low-curvature regions given by $\eta\rightarrow\infty$ and $\eta\rightarrow-\infty$, respectively). By inspecting the mode equation (\ref{modeeq}) we see that its high-curvature ($\eta\rightarrow \infty$) set of solutions is given by the mode functions
\be
\label{invac}
u_k = A J_{-ikr}(x),
\ee
where $J_{-ikr}$ is a Bessel function of the first kind, $x\equiv r\sqrt{\bar{\Lambda}} e^{-\eta/r}$ and $A$ is some normalization constant. Thus we will assume that the high-curvature vacuum $|_{(in)}0\rangle$ (in-vacuum) is completely specified by the set of mode functions given by (\ref{invac}) and is formally given by the following expression:
\be
\label{invacrep}
|_{(in)}0\rangle\equiv\prod_{\vec{k}\in upper} |_{(in)}0_{\vec{k}}\rangle \otimes |_{(in)} 0_{\vec{-k}}\rangle,
\ee
where all $|_{(in)}0_{\vec{k}}\rangle$ are annihilated by the annihilation operators $\hat{a}^{-}_{\vec{k}}$ associated with the mode functions $(\ref{invac})$ while the product goes over all $\vec{k}$ that ends in the upper half-plane defined by axes $k_1$ and $k_2$.

The low-curvature ($\eta\rightarrow -\infty$) set of solutions of (\ref{modeeq}) is given by the mode functions
\be
\label{outvac}
v_{\vec{k}}= B H^{(2)}_{-ikr}(x),
\ee
where $H^{(2)}_{-ikr}$ is a Hankel function of the second kind and $B$ is some normalization constant. We will assume then that the low-curvature vacuum $|_{(out)}0\rangle$ (out-vacuum) is completely specified by the set of mode functions given by (\ref{outvac}) and is formally given by the following expression:
\be
\label{outvacrep}
|_{(out)}0\rangle\equiv\prod_{\vec{k}\in upper} |_{(out)}0_{\vec{k}}\rangle \otimes |_{(out)} 0_{\vec{-k}}\rangle,
\ee
where all $|_{(out)}0_{\vec{k}}\rangle$ are annihilated by the annihilation operators $\hat{a}^{-}_{\vec{k}}$ associated with the mode functions $(\ref{outvac})$ and the product goes over all $\vec{k}$ that ends in the upper half-plane defined by axes $k_1$ and $k_2$.

We also notice that $u_k \sim J_{-ikr}(x)$, asymptotically for $\eta\rightarrow \infty$, are the eigenvectors of $\hat{\pi}_\eta$ to the eigenvalues $\sqrt{\bar{\Lambda}}$ (parameterized by $\bar{\Lambda}$ with $k=\sqrt{\bar{\Lambda}}$). Thus in the high-curvature limit the modes $u_k$ correspond to the collapsing pre-big-bang universe. We also recognize that $v_{k} \sim H^{(2)}_{-ikr}(x)$ and $v_{k}^{*}\sim \left\{H^{(2)}_{-ikr}(x)\right\}^{*}=H^{(1)}_{ikr}(x)$, asymptotically for $\eta\rightarrow -\infty$,
are the eigenvectors of $\hat{\pi}_\eta$ to the eigenvalues  $\sqrt{\bar{\Lambda}}e^{-\frac{\eta}{r}}$ and $-\sqrt{\bar{\Lambda}}e^{-\frac{\eta}{r}}$, respectively. Thus in the low-curvature limit the modes $v_k$ correspond to the collapsing pre-big-bang universe while the modes $v_{k}^{*}$ correspond to the expanding post-big-bang universe.

The following remarks are in order. The hamiltonian given by (\ref{3ham}) explicitly depends on the time variable $\eta$ and thus does not possess well-defined ground state. However, it is still possible to define the so called \emph{instantaneous lowest-energy state of the hamiltonian} (\ref{3ham}) which is defined as a ground state of the hamiltonian for a particular value of the time parameter $\eta$. It can be shown \cite{Mukhanov} that mode functions $v_k(\eta)$ that fulfill the following initial conditions:
\begin{eqnarray}
\label{convac} \nonumber
v_{k}(\eta_0)&=&\frac{1}{\sqrt{\omega(\eta_0)}}, \\
v'_{k}(\eta_0)&=&i\omega(\eta_0)v_{k}(\eta_0),
\end{eqnarray}
for some particular value of the time parameter $\eta_0$, defines a vacuum which is identical with the instantaneous lowest-energy state of the hamiltonian (\ref{3ham}) at $\eta=\eta_0$. Moreover the hamiltonian (\ref{3ham}) at $\eta=\eta_0$ is related to the operators $\hat{a}^\pm_{\vec{k}}$ by:
\begin{eqnarray}
\label{3hamred}
H_{3Q}(\eta)\Bigr\rvert_{\eta= \eta_0}=\int d^2k \omega_k(\eta) \Bigr\rvert_{\eta= \eta_0} \left[ \hat{a}^+_{\vec{k}} \hat{a}^-_{\vec{k}} +\frac{1}{2}\delta^{(2)}(0)  \right],
\end{eqnarray}
which is diagonal in the eigenbasis of the number operator $\hat{N}_{\vec{k}} = \hat{a}^+_{\vec{k}} \hat{a}^-_{\vec{k}}$. Accordingly, the vacuum given by the conditions (\ref{convac}) is sometimes called \emph{the vacuum of instantaneous diagonalization} \cite{Mukhanov}.
In our scenario the mode functions $v_{k}$ given by (\ref{outvac}) fulfill the conditions (\ref{convac}) for $\eta=\eta_0\rightarrow-\infty$ which means that the hamiltonian (\ref{3ham}) in the low-curvature limit (at $\eta=\eta_0\rightarrow-\infty$) reduces to (\ref{3hamred}), where the creation and annihilation operators $\hat{a}^+_{\vec{k}}$ and  $\hat{a}^-_{\vec{k}}$ correspond to the mode functions (\ref{outvac}). We also recognize that the low-curvature vacuum (\ref{outvacrep}) is identical with the lowest-energy state of the hamiltonian (\ref{3ham}) at times $\eta\rightarrow-\infty$.

Let us now specify the boundary conditions related with the considered problem. We will assume that initially for $\eta\rightarrow\infty$ the quantum state of the multiverse $|_{(in)}\Psi\rangle$ is identical with the high-curvature vacuum $|_{(in)}0\rangle$ completely specified by the mode functions (\ref{invac}):
\be
\label{instat}
|_{(in)}\Psi\rangle=|_{(in)}0\rangle.
\ee
In other words, the multiverse is initially (for $\eta\rightarrow\infty$) in a vacuum state. Since the evolution of our setup is formally equivalent to the stationary scattering, the state of the universe does not change and is given by (\ref{instat}) during the whole process. So:
\be
|_{(out)}\Psi\rangle=|_{(in)}\Psi\rangle,
\ee
where  $|_{(out)}\Psi\rangle$ represents the final state of the multiverse for $\eta\rightarrow-\infty$. The state of the vacuum, however, does evolve, since, according to our previously made assumption, it is controlled by the instantaneous value of the curvature. It means that the vacuum state transforms during the whole process form the high-curvature vacuum $|_{(in)}0\rangle$ (in-vacuum for $\eta\rightarrow\infty$) given by (\ref{invacrep}) into the low-curvature vacuum $|_{(out)}0\rangle$ (out-vacuum for $\eta\rightarrow-\infty$)  given by (\ref{outvacrep}). Thus, finally for $\eta\rightarrow-\infty$, the state of the multiverse is not anymore a vacuum state.
For each mode ${\vec{k}}$ we have \cite{Mukhanov}:
\begin{eqnarray}
\label{inout} \nonumber
|_{(in)}0_{\vec{k}}\rangle &\otimes& |_{(in)} 0_{\vec{-k}}\rangle=\\
&=&\frac{1}{|\alpha_{k}|}\sum_{n=0}^{\infty}\left(-\frac{\beta_{k}^*}{\alpha_{k}}\right)^n |_{(out)}n_{\vec{k}}\rangle \otimes |_{(out)}n_{\vec{-k}}\rangle,
\end{eqnarray}
where
\begin{eqnarray}
\label{excstates}
|_{(out)}n_{\vec{k}}\rangle\equiv\frac{1}{\sqrt{n!}}(\hat{a}^+_{\vec{k}})^n|_{(out)}0_{\vec{k}}\rangle,
\end{eqnarray}
with $\hat{a}^+_{\vec{k}}$ being the creation operators associated with the mode functions (\ref{outvac})
while $\alpha_k$ and $\beta_k$ are the Bogolyubov coefficients given by:
\begin{eqnarray}
\label{Bogcoef1}
\alpha_k&=&\frac{W(u_k,v^*_k)}{2i}, \\
\label{Bogcoef2}
\beta_k&=&\frac{W(v_k,u_k)}{2i}.
\end{eqnarray}
In the usual picture of the second quantized Klein-Gordon field, the states given by (\ref{excstates}) are assumed to represent n particles with momentum $\vec{k}$. However, since the third quantization of the Wheeler-DeWitt wave function goes beyond the ordinary scheme of the quantum field theory, sticking to such a standard interpretation seems to be not the only possible option. It is formally viable to interpret vectors given by (\ref{excstates}) as referring to the internal degrees of freedom of some physical setup which exists in the minisuperspace, that is characterized by the momentum $\vec{k}$. To be more precise, we \emph{postulate} that the vacuum states such as $|_{(out)}0_{\vec{k}}\rangle$ refer to a particle (or equivalently a universe) with momentum $\vec{k}$, whose internal state is given by the ground state (the lowest energy state). Consequently, the states given by (\ref{excstates}) refer to a particle with momentum $\vec{k}$, whose internal state is described by the n-th excited state with respect to the ground state given by $|_{(out)}0_{\vec{k}}\rangle$. 
Let us also notice that the states $|_{(out)}n_{\vec{-k}}\rangle$ correspond to the collapsing pre-big-bang branch of the solution given by (\ref{rozwio1}), (\ref{rozwio2}) and (\ref{conect}) while the states $|_{(out)}n_{\vec{k}}\rangle$ correspond to the expanding post-big-bang branch of the mentioned solution. The problem of attributing the energy to the postulated internal states enumerated with the quantum number n, as well as the problem of its interpretation will be tackled in the next section of the paper.

From the formula (\ref{inout}) we see that for $\eta\rightarrow\infty$, which corresponds to the high-curvature limit, the multiverse is composed of pairs of disentangled universes with opposite momenta $\vec{-k}$ and $\vec{k}$ since the quantum state of each such pair is given by $|_{(in)}0_{\vec{k}}\rangle \otimes |_{(in)} 0_{\vec{-k}}\rangle$ which is manifestly a separable state. On the other hand, as $\eta\rightarrow-\infty$, which corresponds to the low-curvature limit, the multiverse transforms into a set of pairs of entangled universes with opposite momenta $\vec{-k}$ and $\vec{k}$ since the right-hand side of the formula (\ref{inout}) represents an entangled state. Thus, the considered scenario results in the emergence of the entanglement in pairs of the universes where each such pair consists of the contracting pre-big-bang and the expanding post-big-bang universe. Following \cite{Robles_ent1,Robles_ent2,Robles_Bal} we will call such an entangled pair of universes \emph{a doubleverse}.

In this paper we will assume the perspective of an observer associated with the expanding branch for which the contracting branch is inaccessible. From his point of view the the state of the expanding branch being a subset of the composite quantum mechanical system made up of both the expanding and the contracting branches is given by the reduced density matrix which is a result of tracing away the degrees of freedom associated with the contracting branch:
\begin{eqnarray}
\label{reduced}
\rho_{\vec{k}}=\sum_{m=0}^{\infty} \langle _{(out)}m_{\vec{-k}}|\rho|_{(out)}m_{\vec{-k}}\rangle,
\end{eqnarray}
where
\begin{eqnarray}
\label{pure}
\rho=|_{(in)}0_{\vec{k}}\rangle \otimes |_{(in)} 0_{\vec{-k}}\rangle \langle_{(in)}0_{\vec{k}}| \otimes \langle_{(in)} 0_{\vec{-k}}|.
\end{eqnarray}
By performing the trace in (\ref{reduced}) we obtain:
\begin{eqnarray}
\label{reducedExpl}
\rho_{\vec{k}}=\frac{1}{|\alpha_{k}|^2} \sum_{m=0}^{\infty} \left| \frac{\beta_k}{\alpha_k} \right|^{2m} |_{(out)}m_{\vec{k}}\rangle \langle _{(out)}m_{\vec{k}}|.
\end{eqnarray}
By normalizing the mode functions $u_k$ and $v_k$ with condition (\ref{normcon}) and then by calculating the wronskians (\ref{Bogcoef1}) and (\ref{Bogcoef2}) we obtain the Bogolyubov coefficients $\alpha_k$ and $\beta_k$ in the following form:
\begin{eqnarray}
\label{BogcoefExpl}
\alpha_k&=&\frac{1}{\sqrt{1-e^{-2 \pi kr }}}, \\
\beta_k&=&\frac{1}{\sqrt{e^{2 \pi k r}-1}}.
\end{eqnarray}
The eigenvalues of the reduced density matrix $\rho_{\vec{k}}$ given by (\ref{reducedExpl}) are:
\begin{eqnarray}
\label{eigenval}
\lambda_n\equiv \frac{1}{|\alpha_{k}|^2} \left| \frac{\beta_k}{\alpha_k} \right|^{2n}=\frac{e^{-2 \pi krn}}{1-e^{-2 \pi kr}}.
\end{eqnarray}
The eigenvalues (\ref{eigenval}) do not fulfill the normalization condition since
\begin{equation}
\label{nomralization}
\sum_{n=0}^{\infty}\lambda_n= \frac{1}{(1-e^{-2 \pi kr })^2}.
\end{equation}
The corrected eigenvalues which fulfill the normalization condition are then:
\begin{eqnarray}
\label{eigenvalcorr}
\tilde{\lambda}_n\equiv(1-e^{-2 \pi kr}) e^{-2 \pi krn}.
\end{eqnarray}

\section{The energy and the entropy of the entangled pair of universes}
\label{sec:3}
\begin{figure}
\begin{center}
\resizebox{0.4\textwidth}{!}{\includegraphics{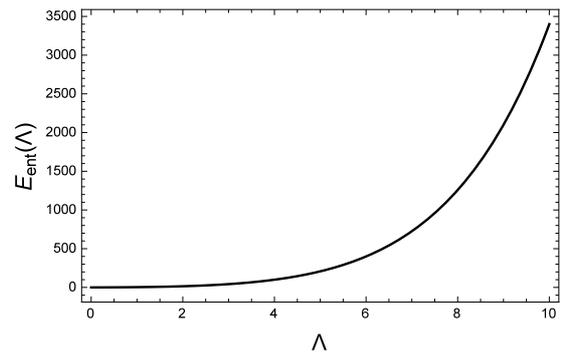}}
\caption{The energy of entanglement $E_{ent}$ against the cosmological constant $\Lambda$. The quantity $E_{ent}$ is a monotonically growing function of the cosmological constant $\Lambda$ and it vanishes for $\Lambda=0$.}
\label{Eent1}
\end{center}
\end{figure}

We will show that with the process of the creation of the pair of entangled universes there is associated a production of the entropy and the energy of entanglement. In order to see that we will first calculate the energy of entanglement (an analog of the internal energy in thermodynamics) defined as \cite{Robles_ent1,Robles_ent2,Robles_Bal}:
\begin{eqnarray}
\label{energy}
E_{ent}\equiv   Tr \left(\rho_{\vec{k}} H_d \right) =   \sum_{n=0}^{\infty}\langle _{(out)}n_{\vec{k}}|\rho_{\vec{k}} H_d |_{(out)}n_{\vec{k}}\rangle,
\end{eqnarray}
where
\begin{eqnarray}
\label{hamdouble}
H_d \equiv  \omega_k(\eta) \left[ \hat{a}^+_{\vec{k}} \hat{a}^-_{\vec{k}} +\frac{1}{2} \right],
\end{eqnarray}
is a hamiltonian of a single universe of the doubleverse. The explicit form of the energy of entanglement is:
\begin{eqnarray}
\label{energyExpl}
E_{ent}=  \frac{\sqrt{\Lambda}}{2}\left(1-x^2\right) e^{-\frac{\eta}{r}},
\end{eqnarray}
where $x$ is defined by
\begin{eqnarray}
\label{x}
x\equiv \left(\frac{\beta_k}{\alpha_k}\right)^2= 
e^{-2\pi kr},
\end{eqnarray}
where $kr= 2 \sqrt{\frac{\tilde{V_0}\bar{\Lambda}}{A^2-9}}$. The energy of the entanglement $E_{ent}$ (see Fig. \ref{Eent1}) grows monotonically together with the value of the cosmological constant $\Lambda$ and it reaches zero as the cosmological constant $\Lambda$ vanishes (we set $\tilde{V_0}=1$).

\begin{figure}
\begin{center}
\resizebox{0.4\textwidth}{!}{\includegraphics{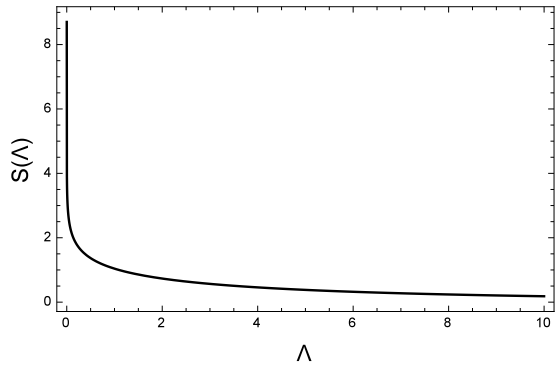}}
\caption{The entropy of entanglement $S(\rho_{\vec{k}})$ against the cosmological constant $\Lambda$. The quantity $S(\rho_{\vec{k}})$ is a monotonically decreasing function of the cosmological constant $\Lambda$ and it reaches infinity for $\Lambda=0$ while goes to zero as $\Lambda$ grows.}
\label{Sent1}
\end{center}
\end{figure}

The entropy of entanglement is given by the von Neumann entropy and is defined as \cite{Robles_ent1,Robles_ent2,Robles_Bal}:
\begin{eqnarray}
\label{entropy}
S(\rho_{\vec{k}})\equiv-\sum_{n=0}^{\infty}\tilde{ \lambda}_n \ln{\tilde{\lambda}_n}.
\end{eqnarray}
By substituting the corrected eigenvalues  $\tilde{\lambda}_n$ of $\rho_{\vec{k}}$  given by (\ref{eigenvalcorr}) into (\ref{entropy}) we obtain that:
\begin{eqnarray}
\label{entropy2}
S(\rho_{\vec{k}})=\ln{\left[\frac{x^{\left(\frac{x}{x-1}\right)}}{1-x}\right]}.
\end{eqnarray}
The entropy of entanglement $S(\rho_{\vec{k}})$ (see Fig. \ref{Sent1})  monotonically decreases as the cosmological constant grows. It becomes infinite for the vanishing cosmological constant while tends to zero as the cosmological constant approaches infinity. In other words the pairs of the universes characterized by small values of the vacuum energy are initially much more entangled than those with larger values of the vacuum energy. In fact if the vacuum energy is very large the entanglement disappears and the state of the pair of the universes becomes separable. On the other hand vanishing of the vacuum energy is accompanied by maximal (infinite) entanglement.
It seems strange that the energy of entanglement $E_{ent}$ (compare  Fig. \ref{Eent1} and Fig. \ref{Sent1}) on one hand vanishes as the entropy of entanglement $S(\rho_{\vec{k}})$ approaches infinity (maximal entanglement) while on the other hand it goes to infinity as the entropy of  entanglement $S(\rho_{\vec{k}})$ approaches zero value. This means that the energy of entanglement is not a good measure of the strength of entanglement. Remember, however, that the quantum number $n$ enumerates the internal excitation levels of the single universe of the considered doubleverse and the quantity $E_{ent}$ defined by (\ref{energy}) gives the average value of the energy associated with the internal excitation levels. Given the above, it seems sensible to think of the energy of entanglement as of the quantity which is associated with a single universe and whose presence should at least in principle be detectable. In the next section we will \emph{postulate} that the energy of entanglement $E_{ent}$ can be noticed by an observer inside a single universe of the doubleverse as the energy which supplements the energy associated with the matter content.

\section{Entanglement effective perfect fluid}
\label{sec:4}
\begin{figure}
\begin{center}
\resizebox{0.4\textwidth}{!}{\includegraphics{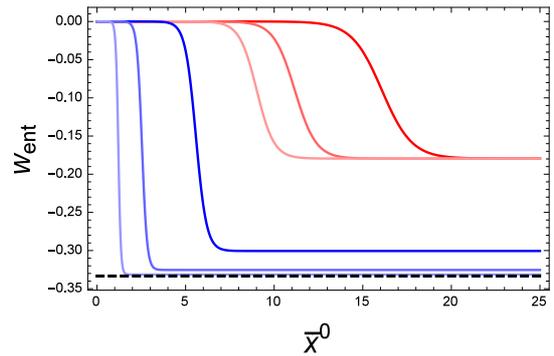}}
\caption{The effective barotropic index $w_{eff}$ against the rescaled proper time of the comoving observer $\bar{x}^0$. Red lines represent models which differs with $\Lambda$ only. Blue lines represent models which differs with $A$ parameter only. All models with $F=0$ are represented by the dashed black line in the above figure (regardless the value of other parameters).}
\label{EoS}
\end{center}
\end{figure}

Assuming that the energy of entanglement can be a part of the energy-momentum content of the single universe the effect of quantum entanglement can manifest itself in the from of the effective prefect fluid which may affect the evolution of the classical background. We additionally assume that the effective fluid does not interact with the other perfect fluid filling the space. In order to derive the form of the associated barotropic index we start with the ordinary continuity equation:
\begin{equation}
\label{continuity}
d\rho+3 \frac{da}{a}(1+w_{ent})\rho=0,
\end{equation}
where $w_{ent}$ is the barotropic index of the effective fluid associated with the effect of the entanglement. The energy density of the effective fluid scales in the following way:
\be
\label{rhoscal}
\rho\sim \frac{E_{ent}}{a^3}.
\ee
Taking into account the expression (\ref{energyExpl}) we can easily calculate that:
\be
\label{epoa}
\frac{dE_{ent}}{da}=-E_{ent}\frac{\frac{dI}{dx^0}}{\frac{da}{dx^0}},
\ee
where $a$ is the scale factor and $I\equiv \frac{\eta}{r}$. By combining  (\ref{continuity}), (\ref{rhoscal}) and (\ref{epoa})  we obtain the effective barotropic index $w_{ent}$ in the following form:
\be
\label{went}
w_{ent}=\frac{a}{3}\frac{\frac{dI}{dx^0}}{\frac{da}{dx^0}}.
\ee
Calculating the derivatives in the equation above  allows us to plot the effective barotropic index $w_{eff}$ against the rescaled proper time of the comoving observer $\bar{x}^0$ for different values of the model parameters (see Fig. \ref{EoS}).
In each case the effective barotropic index $w_{ent}$ suddenly changes it value from zero to a value  between $-0.17$ and $-1/3$.
For higher values of the cosmological constant the transition occurs  earlier and the slope is steeper. Similarly higher value of the $A$ parameter makes that the transition occurs earlier and the slope is steeper. It also results in more negative value of the effective barotropic index $w_{ent}$ after the transition. The analysis of the formula (\ref{went}) shows that for higher value of the kinetic energy related with the free degrees of freedom (determined by the value of the constant $F$) the transition occurs later. On the other hand for sufficiently small value of $F$ the transition disappears and the effective barotropic index maintains a constant value equal to approximately $-1/3$ all the time.
An interesting issue here is the effect of an entanglement backreaction which according to (\ref{energyExpl}) introduces the following correction to the value of the cosmological constant:
\be
\label{backreac}
\Lambda\rightarrow\Lambda(1-x^2)^2.
\ee
By equation (\ref{x}) and Fig. (\ref{Sent1}) we see that the strong entanglement (for high value of the entropy of entanglement) can largely suppress the value of cosmological constant. On the other hand if the entanglement is weak the effect of backreaction disappears. Interestingly the backreaction on the vacuum energy does not affect the classical orbits of the system in the minisuperspace (eq. (\ref{rozwio1}) and (\ref{rozwio2})). However, the backreaction of the entanglement induced effective perfect fluid has to be taken into account since for the case with $w_{ent}=-1/3$  (which corresponds to the cosmic strings) the density of the effective fluid may dominate the vacuum energy at early times.

\section{Conclusions}
\label{sec:conc}
We have shown that the canonical quantization of the Wheeler-DeWitt wave function for non-minimally coupled varying constants model introduced in \cite{Balcerzak1} results in a theory which comprises a scenario that describes the two quantum mechanically entangled - one expanding and one contracting - branches. This is different form the scenario developed in \cite{Balcerzak2} where the third quantization applied to the same model led to a scenario in which a whole multiverse subjected to Bose-Einstein distribution emerged form nothing. The discrepancy in these two scenarios follows form different interpretations of the representation dependent sets of vectors spanning the Hilbert space resulting form the third quantization procedure assumed in both approaches. In scenario given in \cite{Balcerzak2} the orthonormal basis that generates the Hilbert space of the multiverse is assumed to represent an occupation with universes in a given state while in the scenario considered in the present paper the basis that spans the Hilbert space is assumed to represent an excitation levels of one of the two systems which naturally leads to entanglement in a pair of single universes that form the doubleverse (compare with approaches introduced in \cite{Robles_ent1,Robles_ent2,Robles_Bal,R1,R2,R3}). Such an approach also facilitates a description of the entanglement in terms of quantities which are formal analogs \cite{Alicki} of the ordinary thermodynamical quantities such as the entropy, the internal energy, heat and work. Including these analogs in the considerations about the multiverse has for the first time been done in \cite{Robles_ent2}, however, their relation with the ordinary thermodynamical quantities has never been clearly articulated. This seems to be important since any such relation could possibly equip our models with  traits indicating existence of interuniversal entanglement. The postulated relation presented in this paper involves interpreting the energy of entanglement as a form of non-interacting energy homogeneously filling the space. In the framework of our model such assumption results in appearance of perfect fluid with the time dependent barotropic index which may influence the early-time evolution. It should be stressed that our postulate is of a very speculative nature since it was not derived from fundamental principles. However, making such additional assumptions seems to be unavoidable for the interuniversal entanglement to affect in any way the internal properties of a single universe and to become this way an observationally testable phenomenon (compare with the approaches postulating quadratic terms representing an interaction between the universes in the minisuperspace \cite{Serrano,Robles1,Bertolami}).

\end{document}